\documentclass[12pt,a4paper]{article}
\usepackage{amssymb,amssymb,amsmath,amsthm}
\usepackage{graphicx}
\usepackage[cp1251]{inputenc}
\newcommand{\ignore}[1]{}

\newcommand{\BB}{\mathbb B}
\renewcommand{\SS}{\mathbb S}
\newcommand{\B}{\mathcal{B}}
\newcommand{\C}{\mathsf{C}}
\newcommand{\CL}{\mathsf{C}_{\log}}
\newcommand{\ov}{\overline}
\newcommand{\comp}{\circ}
\newcommand{\e}{*}
\newcommand{\ce}{\star}
\newcommand{\lb}{[\mkern-3mu[}
\newcommand{\rb}{]\mkern-3mu]}
\newcommand{\conc}{\,\|\,}
\newcommand{\grad}{\bigtriangledown}
\newcommand{\bin}{\text{bin}}

\newcommand{\add}{\text{\texttt{ADD}}}
\newcommand{\car}{\text{\texttt{CAR}}}
\newcommand{\cmp}{\text{\texttt{CMP}}}
\newcommand{\cyc}{\text{\texttt{CYC}}}
\newcommand{\sft}{\text{\texttt{SFT}}}
\newcommand{\dec}{\text{\texttt{DEC}}}
\newcommand{\enc}{\text{\texttt{ENC}}}

\newcommand{\fod}{\text{\texttt{FOI}}}

\newcommand{\inc}{\text{\texttt{INC}}}
\newcommand{\mmax}{\text{\texttt{MAX}}}

\newcommand{\mux}{\text{\texttt{MUX}}}
\newcommand{\pref}{\text{\texttt{PREF}}}
\newcommand{\prs}{\text{\texttt{PS}}}

\newcommand{\udc}{\text{\texttt{UDC}}}

\newcommand{\grc}{\text{\texttt{GRC}}}

\begin{document}

\title{Complexity of basic boolean operators for digital circuit design\footnote{Translated
 from the Russian original published in: {\it Intellektual`nye sistemy. Teoriya i prilozheniya $[$Intellectual systems. Theory and applications$]$.} 2026. {\bf30}(1), 164--186.}}
\author{Igor~S.~Sergeev\footnote{e-mail: isserg@gmail.com}}
\date{}

\maketitle

\begin{abstract}
This article provides a survey of circuit complexity bounds for
basic boolean transforms exploited in digital circuit design and
efficient methods for synthesizing such circuits. The exposition
covers structurally simple functions and operators, such as
counters, adders, encoders, and multiplexors, and excludes more
complex algebraic operations with numbers, polynomials, and
matrices. Several applications to implementing more specific
operations are also discussed.
\end{abstract}

{\bf \qquad Introduction}

\medskip

This paper attempts to collect information on the complexity of
the simplest and most fundamental boolean transforms, which are
widely used in both practical circuit design and theoretical
circuit synthesis problems. These include encoders, multiplexors,
comparators, and other operators that are structurally no more
complex than addition. These transforms often lack a clear
independent meaning, but rather serve as ``building blocks'' for
solving substantive problems. These transforms are typically the
starting point for studying the fundamentals of digital circuit
design in popular textbooks; see, e.g.,~\cite{hh17,ugr10}.
Therefore, in this survey, we (informally) refer to them as {\it
basic}.

The theory of algorithms for multiplying numbers or matrices, or
discrete Fourier transforms, is so developed that it requires
multi-volume editions to cover. Meanwhile, the ``workhorses'' of
electronics, such as the shift operator, the priority encoder, or
the unary converter, are usually overlooked. Popular monographs on
computational complexity, such as~\cite{knu2,knu4,weg87}, provide
only fragmentary information on the basic operators. The goal of
this survey is to present as complete a picture as possible.

A significant portion of the results and methods considered below
should be relegated to folklore due to their simplicity and
familiarity. Therefore, they are presented here without citing any
references. In other cases, where results requiring considerable
effort are discussed, especially for lower bounds, references to
works known to the author are provided, as is customary. The
author had to fill in some gaps himself, but without resorting to
nontrivial constructions.

We consider the implementation of boolean transforms in the model
of circuits of functional elements, i.e. boolean or logical
circuits, see, e.g.,~\cite{cha12,weg87}. Of the standard
mathematical models, it most closely corresponds to real
electronic circuits. Recall that a {\it circuit over a basis} (a
set of functions)~$\B$ is an acyclic directed graph in which
vertices that have no incoming edges are marked as inputs, and
some vertices are marked as outputs. The inputs are labeled by
symbols of variables or constants of the basis~$\B$, while the
remaining vertices (these vertices are called functional elements
or gates) by symbols of functions from the basis~$\B$. The
functioning of the circuit is defined naturally, from inputs to
outputs: at each vertex, the associated function is computed, the
arguments of which are the functions arriving along the edges
entering the vertex. A circuit implements an operator (a system of
functions)~$F$ if all components of the operator are computed at
the circuit's outputs. The {\it complexity} of a circuit is
defined as the number of vertices in its graph, excluding inputs.
The complexity of an operator~$F$ when implemented by circuits
over a basis~$\B$ is defined as the complexity of the minimal
circuit implementing it. The {\it depth} of a circuit is the
length (measured in edges or functional elements) of the longest
directed input-output path. Similarly, the depth of an
operator~$F$ is defined as the minimal depth of a circuit
implementing it. Actually, the complexity roughly corresponds to
the area of a circuit, and the depth to the propagation delay of a
signal from the inputs to the outputs.

We restrict our consideration to the basis of all binary boolean
functions. We denote the complexity of a circuit $\Phi$ over this
basis by $\C(\Phi)$, and the complexity of an operator~$F$ by
$\C(F)$. We also introduce the functional $\CL(F_n)$, which
denotes the complexity of implementing a sequence of operators
$F_n$ by circuits of depth $O(\log n)$, where $n$ is the number of
input variables. Informally, this is the complexity of parallel
computation of the operator. In practical circuit design, parallel
circuits are preferred. It is worth noting that the operators
considered below are quite simple and can be implemented by
parallel circuits.

When designing electronic circuits, wider bases are typically
available, which may include multi-input gates and even gates with
multiple outputs. However, efficient synthesis methods are
generally quite universal and can be adopted to any basis. Recall
that the order of complexity/depth of a function over any complete
finite basis is the same.

The complexity bounds here characterize the computational
complexity in the asymptotic sense, i.e., for the number of inputs
$n \to \infty$. However, it is well known that asymptotically
efficient synthesis methods do not necessarily yield good results
for practically significant values $n$. However, the simple
methods discussed below usually perform well even for very small
values $n$. Moreover, for such values, parallel synthesis methods
lead to circuits with compromised depth and complexity
characteristics.

Further, we present information on the complexity and efficient
implementation methods of basic transforms. Basic operations
include: prefix and suffix sums, numeric increment/decrement,
up-down counter, Gray counter, carry computation, addition and
comparison of two numbers, maximum/minimum of two numbers,
decoder, multiplexor, direct and cyclic shift, encoder, extraction
of the first one and its position number, bit summation and
comparison of the sum with a threshold, computation of the width
of a block of ones, conversion between binary and unary encodings,
truncating, and sorting array of bits. Complexity bounds for these
operations are summarized in Table~\ref{t1}.

The survey is supplemented with examples of application of basic
operations and some useful design ideas to constructing parallel
circuits for a two-selector, a weight-preserving counter, multiple
selection, and permutation of a pair of bits.

The following notations are used throughout the presentation:

$\BB = \{0,\,1\}$;

$\BB^n$ --- the set of boolean strings or vectors of length $n$;
by default, the bits in a string are numbered from zero, left to
right;

$\lb n \rb$ --- the set of integers from $0$ to $n-1$, specified
in binary notation of length $\lceil \log n \rceil$, the bits are
numbered from zero, right to left;

$|s|$ --- the length of a boolean vector or string;

$\nu(X)$ --- the binary weight (the number of ones) of a boolean
string or number $X$;

$\ov x$, $x \vee y$, $x \wedge y$ or $x\cdot y$, $x \oplus y$, $x
\sim y$ --- boolean operations of negation, disjunction,
conjunction, addition modulo 2, and equivalence;

$X^{\alpha} = \bigwedge x_i^{\alpha_i}$ --- elementary conjunction
of a vector of variables $X = [x_1,\, x_2,\, \ldots]$, where
$\alpha = [\alpha_1,\, \alpha_2,\, \ldots] \in \BB^{|X|}$; by
definition, $x^1=x$ and $x^0=\ov x$;

``$\conc$'' --- string concatenation operation;

$[\sigma]^m$ --- a vector or string of length $m$ consisting of
boolean values $\sigma$.

All logarithms below are base 2.

\medskip

{\bf \qquad Complexity of basic operators}

\medskip

{\bf Prefix sums.} The operator $\pref^{\e}_n: \SS^{n} \to
\SS^{n}$ computes a family of prefix sums of $n$~variables from
the semigroup $(\SS,\,\e)$:
\begin{equation}\label{pref}
p_i = x_1 \e x_2 \e \ldots \e x_{i}, \quad 1\le i \le n.
\end{equation}
Further applications, with the exception of constructions of carry
circuits, will be restricted to the case $\SS=\BB$ and $\e \in \{
\vee, \wedge, \oplus \}$. Circuits that compute the
system~(\ref{pref}) over the basis~$\{\e\}$ are called prefix
circuits.

\begin{figure}[htb]
\begin{center}
\begin{picture}(182,115)(2,-10)


\multiput(30,75)(50,0){4}{\circle*{3}}
\multiput(55,50)(25,0){2}{\circle*{3}}
\multiput(105,25)(25,0){2}{\circle*{3}}
\multiput(180,25)(0,25){2}{\circle*{3}} \put(155,0){\circle*{3}}

\multiput(5,95)(50,0){4}{\line(5,-4){25}}
\multiput(55,95)(25,0){2}{\line(0,-1){45}}
\multiput(105,95)(25,0){4}{\line(0,-1){70}}
\put(30,95){\line(0,-1){20}} \put(155,0){\line(0,1){25}}
\multiput(30,75)(100,0){2}{\line(2,-1){50}}
\multiput(30,75)(50,-25){3}{\line(1,-1){25}}
\put(80,50){\line(2,-1){50}} \put(80,50){\line(4,-1){100}}

\put(2,99){$\scriptstyle x_1$} \put(27,99){$\scriptstyle x_2$}
\put(52,99){$\scriptstyle x_3$} \put(77,99){$\scriptstyle x_4$}
\put(102,99){$\scriptstyle x_5$} \put(127,99){$\scriptstyle x_6$}
\put(152,99){$\scriptstyle x_7$} \put(177,99){$\scriptstyle x_8$}

\put(27,65){$\scriptstyle p_2$} \put(52,40){$\scriptstyle p_3$}
\put(77,40){$\scriptstyle p_4$} \put(102,15){$\scriptstyle p_5$}
\put(127,15){$\scriptstyle p_6$} \put(152,-10){$\scriptstyle p_7$}
\put(177,15){$\scriptstyle p_8$}

\end{picture}
\caption{Ofman's (Brent---Kung) prefix circuit}\label{pic_pref}
\end{center}
\end{figure}
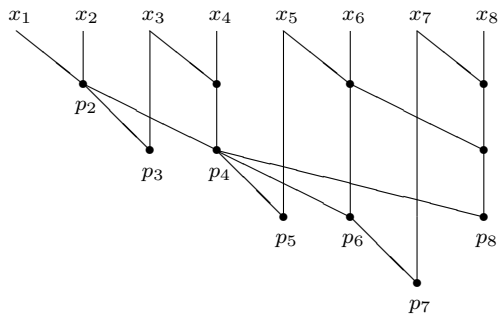

There is an extensive literature devoted to prefix circuits, see,
e.g.,~\cite{ble93,serg13p},\cite[\S3.1]{weg87}. We restrict
ourselves to only brief information. Obviously, the operator
$\pref^{\e}_n$ may be implemented by a circuit of $n-1$
gates~$\e$, in which the prefix sums are calculated sequentially.
The complexity~$C$ and the depth~$D$ of a circuit of
operations~$\e$ that calculates prefix sums of $n$ variables are
related as $C+D \ge 2n-2$. Therefore, the complexity of a parallel
circuit is at least $2n-\Theta(\log n)$. This bound is achieved,
for instance, by circuits proposed by Yu.\,P.~Ofman~\cite{ofm62}
(but in the literature they are usually called Brent---Kung
circuits). In the circuit with $n=2^k$ inputs, the highest sum
$p_n$ is computed by a complete binary tree~$T$ of depth~$k$. Any
missing sum $p_i$ is computed as $p_{\grad i} \e p'$, where $\grad
i$ denotes the number obtained from $i$ by setting the least
significant one to zero, and $p'$ is an appropriate subsum
computed in the tree $T$. It is easy to verify that such circuit
has depth $2k -2$. The 8-input circuit is shown in
Fig.~\ref{pic_pref}. The circuit with an arbitrary number $n <
2^k$ inputs is obtained by truncating the $2^k$- input circuit.

In the case of boolean operations $\e$, all the mentioned bounds
hold for circuits over the binary boolean basis, precisely,
$\C(\pref_n^{\e}) = n-1$ and $\CL(\pref_n^{\e}) = 2n-\Theta(\log
n)$.

There is also the problem of jointly computing prefix and suffix
sums of $n$~variables. The latter are defined as
\begin{equation*}
s_i = x_i \e x_{i+1} \e \ldots \e x_{n}, \quad 1\le i \le n.
\end{equation*}
The complexity of the corresponding operator $\prs^{\e}_n: \SS^{n}
\to \SS^{2n-1}$ is clearly $2n-3$. Minimal parallel prefix-suffix
circuits have complexity $3n-\Theta(\log n)$. Such a circuit can,
for example, be constructed by combining Ofman's prefix and suffix
circuits: the tree computing the sum $p_n=s_1$ of all variables is
the common part for these two circuits.

{\bf Incrementor.} The operator $\inc_n: \lb 2^n \rb \to \lb 2^{n}
\rb$ increments an $n$-bit number by one modulo $2^n$.

The upper bound $\C(\inc_n) \le 2n-2$ for $n \ge 2$ can be easily
obtained. If we denote the input by $X=[x_{n-1}, \ldots, x_0] \in
\lb 2^n \rb$, the digits of the sum $X+1 \bmod 2^n = [z_{n-1},
\ldots, z_0]$ are determined by formulas\footnote{The formula also
holds for $k=0$, if we adopt the convention that the empty
conjunction is equal to~1.}
\begin{equation}\label{inc}
z_k = x_k \oplus x_{k-1} \cdot \ldots \cdot x_0.
\end{equation}
Compute the products in these formulas via the operator
$\pref^{\wedge}_{n-1}$, and then perform bitwise modulo-2 addition
of vectors of length $n$. Employing a parallel prefix circuit
leads to the bound $\CL(\inc_n) \le 3n-\Theta(\log n)$.

The complementary lower bound $\C(\inc_n) \ge 2n-2$ is probably
easier to prove by considering the operator $\inc'_n$, which
computes the sum $X+1$ entirely. It is easy to verify that
$\C(\inc'_n) = \C(\inc_n) + 1$ for $n \ge 2$. Further,
$\C(\inc'_n) = 2n-1$, since substituting $x_{n-1}=0$ into the
circuit for $\inc'_{n}$ reduces the complexity by at least 2, and
the resulting circuit computes the operator $\inc'_{n-1}$.

Decrement circuits that compute the difference $X - 1 \bmod 2^n =
[z_{n-1}, \ldots, z_0]$ may be constructed dually: the digits of
the difference are determined by formulas
\begin{equation}\label{decr}
 z_k = x_k \oplus \ov{x_{k-1}} \cdot \ldots \cdot \ov{x_0}.
\end{equation}

{\bf Up-down counter.} The operator $\udc_n: \lb 2^n \rb \times
\BB \to \lb 2^{n} \rb$, depending on the control input $\sigma \in
\BB$, either increments (for $\sigma=1$) or decrements (for
$\sigma=0$) an $n$-bit number modulo $2^n$.

The bounds $\C(\udc_n) \le 3n-3$ for $n \ge 2$ and $\CL(\udc_n)
\le 4n-\Theta(\log n)$ are obtained by combining the incrementor
and decrementor circuits from the previous paragraph. The result's
digits are computed by formulas combining~(\ref{inc})
and~(\ref{decr}):
\begin{equation*}
z_k = x_k \oplus x_{k-1}^{\sigma} \cdot \ldots \cdot x_0^{\sigma}.
\end{equation*}
Boolean powers $x_k^{\sigma}$ are computed simply as $\sigma \sim
x_k$, $k=0, \ldots, n-2$. Then, the operator
$\pref^{\wedge}_{n-1}$ is applied, followed by bitwise addition of
$n$-bit vectors.

{\bf Gray counter.} The operator $\grc_n: \BB^n \to \BB^{n}$,
given a Boolean string of length $n$, computes the next string
according to the standard Gray encoding in cyclic order. In other
words, it is an incrementor in Gray encoding. In this paragraph,
we number digits of strings from right to left.

Recall that a sequence $G_n$ of $n$-bit Gray strings can be
defined recursively: $G_1 = (0,\,1)$ and then $G_{k+1} = (0\conc
G_k,\, 1\conc G_k^R)$, where $G_k^R$ is the sequence $G_k$ written
in reverse order; the concatenation operation is applied
string-wise. The main property of Gray sequences is that the next
string differs from the previous one in exactly one position. For
more details, see, e.g.,~\cite[Chapter~13]{war03}.

The upper bounds $\C(\grc_n) \le 4n-7$ and $\CL(\grc_n) \le
6n-\Theta(\log n)$ are obtained by conversion to and from the
binary encoding. Let $\bin_n(X)$ denote the binary representation
of the index of a string $X$ in the sequence $G_n$. Then
$\grc_n(X) = \bin_n^{-1}( \inc_n(\bin_n(X)))$. By denoting
$[y_{n-1},\ldots, y_0] = \bin_n(x_{n-1},\ldots, x_0)$, it is easy
to verify that $y_{n-1} = x_{n-1}$ and for any ${i \le n-2}$,
\[ y_i = x_{i} \oplus x_{i+1} \oplus \ldots \oplus x_{n-1}, \qquad x_{i} = y_{i+1} \oplus y_{i}. \]
Therefore, the conversion $\bin_n$ may be performed by the
operator $\pref_n^{\oplus}$, and $\bin_n^{-1}$ may be reduced to
bitwise addition of strings of length $n-1$. It is slightly more
convenient to compute the complement bits $\ov{y_i}$ instead of
$y_i$. When implementing the operator $\inc_n(X)$ in sequential
manner, the two least significant bits do not need to be computed,
since they are equal to $x_0$ and $\ov{y_0}$. The remaining part
of the increment circuit has complexity $2n-4$. Another operation
may be saved in the last step of the conversion to Gray encoding:
the least significant bit of the result is simply $\ov{y_1}$.

{\bf Carry operator.} The operator $\car_n: (\BB^{n})^2 \to
\BB^{n}$ computes the system of carry functions
\begin{equation}\label{car}
c_1 = x_0, \qquad c_{i+1} = x_{i} \oplus y_{i} c_{i}, \quad i = 1,
\ldots, n-1,
\end{equation}
of boolean variables $x_0, x_1, \ldots, x_{n-1}$ and $y_0, y_1,
\ldots, y_{n-1}$.

It follows directly from the definition~(\ref{car}) that
$\C(\car_n) = 2n-2$. The upper bound $\CL(\car_n) \le {5n -
\Theta(\log n)}$ is proved by reducing to the computation of a
system of prefix sums. Define the binary operation $\ce$ on the
set of boolean length-2 vectors as ${[a_1,\, b_1] \ce [a_2,\, b_2]
= [a_2 \oplus b_2a_1,\, b_2b_1]}$. It is easy to verify that the
introduced operation is associative, hence $(\BB^2,\,\ce)$ is a
semigroup. With the notation $p_i = y_{i-1} \cdot \ldots \cdot
y_1$, the system~(\ref{car}) is transformed into a system of
prefix sums:
\begin{equation*}\label{pcar}
[c_1,\,p_1] = [x_0,\,1], \qquad [c_{i+1},\,p_{i+1}] = [c_i,\,p_i]
\ce [x_i,\, y_i], \quad i = 1, \ldots, n-1.
\end{equation*}
Note that the products $p_i$ themselves do not need to be computed
when implementing carries. Then, for the complexity of parallel
carry circuits, we have the bound $\CL(\car_n) \le
{\CL(\pref^{\ce}_n) - (n-1)}$. It remains to observe that the
complexity of the operation $\ce$ is 3.

The idea of computing carries via parallel prefix circuits arose
no later than the 1960s, for example, in~\cite{sk60,ofm62}.

{\bf Adder.} The operator $\add_n: \lb 2^n \rb^2 \to \lb 2^{n+1}
\rb$ calculates the sum of two $n$-bit integers $A$ and $B$.

The upper bounds $\C(\add_n) \le 5n-3$ and $\CL(\add_n) \le 8n-
\Theta(\log n)$ are obtained by attaching two layers of $3n-1$
gates to the carry circuits from the previous
paragraph\footnote{In~\cite{lf80,weg87}, complexity estimates for
parallel adders are given in the form $\CL(\add_n) \le 8n+O(1)$.}.
Denote $A=[a_{n-1}, \ldots, a_0]$, $B=[b_{n-1}, \ldots, b_0]$, and
$A+B=[z_{n}, \ldots, z_0]$. Set $x_i = a_i \wedge b_i$, $y_i = a_i
\oplus b_i$. Let $c_i$ be defined according to~(\ref{car}). Then
$z_0 = y_0$, $z_n=c_n$, and $z_i = y_i \oplus c_i$ for other
$i$'s.

N.\,P. Red'kin~\cite{red81} showed that the former of the two
bounds is tight: in fact, $\C(\add_n) = 5n-3$.

We do not consider the integer subtraction separately, since
negative numbers are usually written in the complement code, in
which subtraction and addition are performed uniformly using
adders, see, e.g.,~\cite[Sec. 4.1]{knu2},~\cite[\S3.1]{weg87}.

{\bf Comparator.} The function $\cmp_n: \lb 2^n \rb^2 \to \BB$
compares $n$-bit numbers $A, B$, i.e., it evaluates the predicate
$A > B$.

The upper bound $\C(\cmp_n) \le 4n-3$ is obtained
straightforwardly. Let $A=[a_{n-1}, \ldots, a_0]$, $B=[b_{n-1},
\ldots, b_0]$, and $x_i = a_i \wedge \ov{b_i}$, $y_i = a_i \sim
b_i$. Then $\cmp_n(A,B) = c_{n}$, where $c_n$ is defined according
to~(\ref{car}). Therefore, $\C(\cmp_n) \le 2n-1 + \C(\car_n)$,
given that there is no need to compute $y_0$.

The bound $\CL(\cmp_n) \le 2n-1 + \CL(c_n) \le 5n-\Theta(\log n)$
is established similarly. Compute $c_{n}$ by a depth-$d$ parallel
prefix tree composed of $n-1$ gates~$\ce$. Note that $d$ gates
computing the second components of the prefix sums (products of
the input variables) can be eliminated.

Non-strict comparison is implemented similarly, since the
predicate $A \ge B$ is the negation of the predicate $B > A$.

The extended operator $\cmp^*_n: \lb 2^n \rb^2 \to \BB^2$
additionally computes the predicate ${A=B}$, i.e., the product
$y_0 \cdot y_1 \cdot \ldots \cdot y_{n-1}$. Its complexity is
estimated as $\CL(\cmp^*_n) \le 5n-3$. Just add the calculation of
$y_0$ and all the second components of the prefix sums into the
parallel comparator circuit described above. The desired product
is finally obtained from these components.

{\bf Maximum of two numbers.} The operator $\mmax_n: \lb 2^n \rb^2
\to \lb 2^n \rb$ computes the maximum of two $n$-bit numbers.

The upper bounds $\C(\mmax_n) \le 6n-3$ and $\CL(\mmax_n) \le
7n-\Theta(\log n)$ are obtained by attaching a layer of $2n$ gates
to the comparator circuits from the previous section.

Denote $c = \cmp_n(A,B)$, then $\max(A, B) = \ov{(A \sim
B)}\cdot[c]^n \oplus B$, where the operations in the last formula
are bitwise. Recall that the vector $A \sim B$, except for its
least significant bit, has already been computed by the comparator
circuit. Therefore, to determine any digit of the result, it is
sufficient to perform two additional operations. Finally, note
that the most significant bit of the maximum is simply $a_{n-1}
\vee b_{n-1}$.

To compute the minimum along with the maximum, it is sufficient to
attach another $n$~gates to the circuit, since $\min(A, B) =
\ov{(A \sim B)}\cdot[c]^n \oplus A$.

{\bf Decoder.} The operator $\dec_n: \lb n \rb \to \BB^{n}$
computes a boolean string of length~$n$ with a~single one at a
given position. The components of the operator are elementary
conjunctions~$X^{\alpha}$ of variables $X$, where the vector
$\alpha$ runs over the set of binary representations of numbers
from~$0$ to $n-1$.

The upper bound $\CL(\dec_n) \le n + \Theta(\sqrt n)$ is obtained
trivially by dividing the set of variables in half: if $X = [X_2,
X_1]$, then $X^{\alpha_2 \conc \alpha_1} = X_2^{\alpha_2} \wedge
X_1^{\alpha_1}$. Hence, $\CL(\dec_n) \le n + \CL(\dec_{2^{k}}) +
\CL(\dec_{\lceil n2^{-k} \rceil})$, where $k = |X_1|$. It remains
to choose $k \approx \log n /2$.

A lower bound of the form $\C(\dec_n) \ge n + \Theta(\sqrt n)$ is
easily established by observing that the set of circuit elements
immediately preceding the outputs has cardinality at least $\sqrt
n$.

The extended operator $\dec^*_n: \lb n \rb \times \BB \to \BB^{n}$
has an additional information input $y\in\BB$ and computes a
vector with bit $y$ in a given position and the rest set to zeros.
Such an operator is often called a demultiplexor. Its complexity
differs from that of the decoder by no more than 2: in the decoder
circuit, it suffices to replace the least significant variable $x$
and its negation with $xy$ and $\ov x \cdot y$, respectively.

{\bf Multiplexors.} The function $\mux_n: \lb n \rb \times \BB^{n}
\to \BB$ selects one of $n$ information boolean variables by its
index (otherwise called an address). This function is called an
$(n,\,1)$-multiplexor, as well as a selector, a switching
function, or a memory access function.

The best known upper bound $\CL(\mux_n) \le 2n+O(\sqrt n)$ was
obtained by P.~Klein and M.~Paterson in~\cite{kp80}.
Represent the address input as $X=[X_2,\,X_1]$, where $|X_1|=q$. 
Divide the string of information variables into blocks of length
$2^{q}$: $Y = Y_0 \conc Y_1 \conc \ldots \conc Y_{p-1}$, $p =
\lceil n/2^{q} \rceil$ (the last block can be shorter). Decompose
the function $\mux_n$ over variables~$X_2$:
\begin{equation}\label{mux0}
\mux_n(X;Y) = \bigvee_{\alpha = 0}^{p-1} X_2^{\alpha}\cdot
\mux_{|Y_{\alpha}|}(X_1;Y_{\alpha}).
\end{equation}
Inner multiplexor functions can be calculated by formulas
\begin{equation}\label{mux1}
\mux_{|Y_{\alpha}|}(X_1;Y_{\alpha}) =
\bigvee_{\beta=0}^{|Y_{\alpha}|-1} y_{\alpha,\beta} \cdot
X_1^{\beta},
\end{equation}
where the introduction of a two-index numbering on the set of
variables~$Y=\{y_{\alpha,\beta}\}$ is implied.

All elementary conjunctions of groups of variables $X_1$ and $X_2$
are computed with complexity of order $2^{q} + p$ using the
decoders $\dec_{2^{q}}$ and $\dec_{p}$. Another $2n + p$
operations are sufficient to complete the computations by
formulas~(\ref{mux0}),~(\ref{mux1}). It remains to choose $q
\approx \log n/2$.

This bound is asymptotically tight: W.~Paul~\cite{paul77}
established that $\C(\mux_n) \ge 2n-2$.

A more general problem often arises: implementing an
$(n,\,k)$-multiplexor $\mux^k_n: {\lb n \rb \times (\BB^k)^{n} \to
\BB^k}$, whose information inputs are $k$-bit numbers or boolean
vectors. The circuit for the operator $\mux^k_n$ is obtained by
parallel combining the $(n,\,1)$-multiplexor circuits described
above for each of the $k$ bits. These circuits have a common part
$\dec_{2^{q}}(X_1)$ and $\dec_{p}(X_2)$. Choosing $q = \lceil
\min(\log n,\,\log(kn)/2)\rceil$ yields the bound $\CL(\mux^k_n)
\le {2kn+O(\sqrt{kn})}$.

{\bf Shift operators.} The operator $\cyc_{k,n}: \lb k \rb \times
\BB^n \to \BB^n$ performs a cyclic shift of a~boolean vector of
length $n$ by $X \in \lb k \rb$ positions\footnote{The direction
of the shift is irrelevant for complexity estimates.}.

The upper bound $\CL(\cyc_{k,n}) \le 3\lceil \log k\rceil n$ is
trivial. The corresponding circuit has the form of an $l$-fold
composition $\mux^n_2 \comp \ldots \comp \mux^n_2$, where $l =
\lceil \log k\rceil$. Write $X = [x_{l-1}, \ldots, x_1, x_0]$. The
first subcircuit performs a cyclic shift by $x_0$, the second by
$2x_1$, the third by $4x_2$, and so on.

The regular shift operator $\sft_{k,n}: \lb k \rb \times \BB^n \to
\BB^{n+k-1}$ is implemented similarly. The bound $\CL(\sft_{k,n})
\le 3\lceil \log k\rceil n - 2(k-1)$ holds, taking into account
that $k-1$ of $(2,\,1)$-multiplexors in the above circuit may be
reduced to a single operation.

{\bf Encoder.} The operator $\enc_n: \BB^{n} \to \lb n \rb$ is a
linear boolean operator with a matrix~$U_n$ of size $\lceil \log n
\rceil \times n$, whose columns contain sequentially the numbers
from $0$ to $n-1$ in binary representation\footnote{For $n=2^k$,
this matrix is the parity-check matrix of the Hamming code up to
the presence of a zero column.}. For an input vector of weight~1,
the encoder determines the position of a~single one. Therefore, on
the set of vectors of weight~1, the encoder $\enc_n$ is the
inverse transform of $\dec_n$.

It is known that $\C(\enc_n) = \CL(\enc_n) = {2(n-\lceil \log
n\rceil -1)}$. The lower bound was proved by
A.~V.~Chashkin~\cite{cha94}. The upper bound can be obtained
trivially via the well-known relation between the complexity of
linear operators with transposed matrices (the transposition
principle~\cite{ms65}): the complexity of an operator with a
matrix~$U_n^{T}$ is simply $n-\lceil \log n\rceil -1$. However, it
is not hard to explicitly describe the construction of the
corresponding circuits. We define a family of circuits $\Phi_n$
computing $\enc_n$ recursively, together with circuits $\Phi'_n$
that implement transforms $\enc'_n$ with matrices $U_n$ padded
with rows of all ones. Let $n = 2^k+p \le 2^{k+1}$. Split a
vector~$X$ of length~$n$ into two parts $X_1, X_2$ of length $2^k$
and $p$. The circuit $\Phi_n(X)$ is obtained from
$\Phi_{2^k}(X_1)$ and $\Phi'_{p}(X_2)$ by attaching additional
$\lceil \log p \rceil$ gates $\oplus$, and $\Phi'_n(X)$ is
obtained from $\Phi'_{2^k}(X_1)$ and $\Phi'_{p}(X_2)$ by adding
$\lceil \log p \rceil+1$ gates. Recurrence relations
\[ \C(\Phi_n) = \C(\Phi_{2^k}) + \C(\Phi'_p) + \lceil \log p \rceil, \qquad \C(\Phi'_n) = \C(\Phi'_{2^k}) + \C(\Phi'_p) + \lceil \log p \rceil + 1\]
are resolved as $\C(\Phi_n) = 2(n-\lceil \log n\rceil -1)$ and
$\C(\Phi'_n) = 2n - \lceil \log n\rceil -2$ taking into account
the initial conditions $\C(\Phi_1) = \C(\Phi'_1) = 0$. The depth
of $\Phi_n$ and $\Phi'_n$ circuits is $\lceil \log n\rceil-1$
and~$\lceil \log n\rceil$ respectively. See~\cite{cha12} for more
details.

In an alternative definition, the encoder $\enc^*_n$ is a partial
boolean operator defined only on the set of weight-1 vectors and
coinciding with $\enc_n$ on this set. Following the proof of the
lower bound in~\cite{cha12}, it is easy to verify that weakening
the definition does not yield a~significant gain: $\C(\enc^*_n) =
2n-\Theta(\log n)$.

{\bf Unary encoding.} The operator $\text{\texttt{UN}}_n:
[\mkern-3mu[ n+1 ]\mkern-3mu] \to \mathbb B^{n}$ converts a number
from standard binary notation to unary notation in which the
number $k$ is written as a string starting with $k$~ones and
followed by zeros.

The upper bound $\mathsf{C}(\text{\texttt{UN}}_n) \le 2n+O(\sqrt
n)$ is easily achieved by a circuit of type
$\text{\texttt{PREF}}^{\vee} \circ \text{\texttt{DEC}}$. The
operator $\text{\texttt{DEC}}_{n+1}(x)$ computes a vector with a
single one at position $x$. Discarding the least significant
component of the vector, compute the suffix sums of the remaining
components.

The given bound may be improved to
$\mathsf{C}_{\log}(\text{\texttt{UN}}_n) \le 2n+O(\sqrt n)$.
First, inductively construct circuits for
$\text{\texttt{UN}}_{2^k-1}$ of complexity $2(2^k-k-1)$ and depth
$k-1$. Let a boolean string $S = [s_1,\ldots,s_{2^k-1}]$ be
obtained as $S = \text{\texttt{UN}}_{2^k-1}(X)$, where $X$ is a
boolean vector representing a $k$-bit number. Then for $y \in
\BB$,
\begin{equation*}
\text{\texttt{UN}}_{2^{k+1}-1}(y,X) = \begin{cases} S \,\|\, [0]^{2^k}, & y = 0 \\
[1]^{2^k} \,\|\, S, & y = 1 \end{cases} \; = [s_1 \vee y,\;
\ldots,\; s_{2^k-1} \vee y,\; y,\; s_1 \cdot y,\; \ldots,\;
s_{2^k-1}\cdot y].
\end{equation*}
Thus, the circuit for $\text{\texttt{UN}}_{2^{k+1}-1}$ is
constructed from the circuit for $\text{\texttt{UN}}_{2^k-1}$ by
adding a layer of $2(2^k-1)$ conjunction and disjunction gates.
Hence, given $\mathsf{C}(\text{\texttt{UN}}_1)=0$, the required
estimates follow.

Next, we construct a circuit for $\text{\texttt{UN}}_n$ using the
block method. Let $X=[X_2,\, X_1]$, where $|X_1| = k \approx \log
n/2$. Denote $p = \lceil (n+1)/2^k \rceil$. The result
$\text{\texttt{UN}}_n(X)$ can be written in block form as $B_0
\,\|\, B_1 \,\|\, \ldots \,\|\, B_{p-1}$, where all substrings
$B_i$ have length~$2^k$, except for the last that is
incomplete\footnote{Its length is from 0 to $2^k-1$.}. Compute
$\text{\texttt{DEC}}_p(X_2) = [a_0, a_1, \ldots, a_{p-1}]$. This
is a vector indicating the position of the first not all-ones
block. Such a block has the form $S =
\text{\texttt{UN}}_{2^{k}-1}(X_1) \,\|\, 0$ (shortened in the case
of the last block). To the left of it the blocks are all-ones, to
the right are all-zeros. Via the operator
$\text{\texttt{PREF}}_p^{\vee}$ compute the prefix sums $b_i =
{a_0 \vee a_1 \vee \ldots \vee a_{i}}$. Then we have
\begin{equation}\label{monB}
B_i = \begin{cases} [1]^{2^k}, & a_i=b_i=0 \\ S, & a_i=b_i=1 \\
[0]^{2^k}, & a_i=0, b_i=1 \end{cases} \; = S \cdot [a_i]^{2^k}
\vee [\overline{b_i}]^{\,2^k},
\end{equation}
where on the right-hand side, operations with boolean strings are
performed bitwise; for $i=p-1$, the block is truncated. The
resulting circuit is composed of subcircuits
$\text{\texttt{DEC}}_p$, $\text{\texttt{PREF}}_p$,
$\text{\texttt{UN}}_{2^{k}-1}$ of complexity $O(\sqrt n)$ and a
layer of $2n$ operations following~(\ref{monB}).

The inverse transform $\text{\texttt{UN}}^{-1}_n: \mathbb B^{n}
\to [\mkern-3mu[ n+1 ]\mkern-3mu]$ from unary to binary encoding
is easy: $\mathsf{C}_{\log}(\text{\texttt{UN}}^{-1}_n) = n-1$. The
digits of the number $[y_k, \ldots, y_0] =
\text{\texttt{UN}}^{-1}_n(s_1, \ldots, s_n)$ are determined by
formulas
\[ y_j = \bigoplus_{i\ge 1} s_{i\cdot2^j}. \]
All the necessary sums can be computed in a tree of $n-1$ gates
$\oplus$ with depth $\lceil \log n \rceil$.

{\bf Truncation.} The operator $\text{\texttt{TRN}}_n:
[\mkern-3mu[ n+1 ]\mkern-3mu] \times \mathbb B^{n} \to \mathbb
B^n$ in a boolean string of length $n$ preserves the first $k$
bits and fills the rest of the string with zeros.

The upper bound $\mathsf{C}_{\log}(\text{\texttt{TRN}}_n) \le
3n+O(\sqrt n)$ is straightforward since $\text{\texttt{TRN}}_n(k;
X) = \text{\texttt{UN}}_n(k) \wedge X$ (the conjunction is
performed bitwise).

{\bf First-one indicator.} The operator $\fod_n: \BB^{n} \to
\BB^{n}\times \BB$ leaves the very first one in a boolean string
of length $n$, replacing the rest with zeros, and additionally
computes the indicator of the presence of a one in the string.

It is easy to see that the operator transforms the string $X =
[x_0, x_1, \ldots, x_{n-1}]$ into $Y = [y_0, y_1, \ldots,
y_{n-1};\, z]$, where $y_k = \ov{(x_0 \vee \ldots \vee
x_{k-1})}\cdot x_k$ and $z = x_0 \vee \ldots \vee x_{n-1}$. To
compute it, it suffices to add a layer of $n-1$ gates of type
$\ov{a}\wedge b$ to a circuit that computes $\pref_n^{\vee}(X)$.
Consequently, $\C(\fod_n) \le 2n-2$ and $\CL(\fod_n) \le
3n-\Theta(\log n)$.

The first bound is tight: it is easy to check that the circuit
implementing $\fod_n$, after substituting $x_{n-1}=0$, computes
$\fod_{n-1}$ and at the same time at least two gates may be
eliminated.

{\bf Priority encoder.} The operator $\text{\texttt{PENC}}_n:
\mathbb B^{n} \to [\mkern-3mu[ n ]\mkern-3mu] \times \mathbb B$
determines the position of the first one in a boolean string of
length $n$ and additionally computes the indicator of the presence
of ones in the string\footnote{This is a partial boolean operator:
the position is undefined for zero input.}.

Upper bounds $\mathsf{C}(\text{\texttt{PENC}}_n) \le 2n-2$ and
$\mathsf{C}_{\log}(\text{\texttt{PENC}}_n) \le 3n-\Theta(\log n)$
are achieved by circuits of the form $\text{\texttt{UN}}^{-1}
\circ \text{\texttt{PREF}}$. Indeed, the first-one position of a
string $X$ can be found as
$\text{\texttt{UN}}_n^{-1}(\overline{\text{\texttt{PREF}}_n^{\vee}(X)})$,
where the negation is applied bitwise. By the way, the internal
operator $\text{\texttt{PREF}}^{\vee}_n$ computes the presence of
ones. For $n \ge 3$, even $\mathsf{C}(\text{\texttt{PENC}}_n) \le
2n-3$ is true: the first-one position is correctly computed as
$\text{\texttt{UN}}_{n-1}^{-1}(\overline{\text{\texttt{PREF}}_{n-1}^{\vee}(X')})$,
where $X'$ is the string~$X$ with the last bit removed.

The lower bound can be stated as
$\mathsf{C}(\text{\texttt{PENC}}_n) \ge
\mathsf{C}(\text{\texttt{ENC}}^*_n) = 2n-\Theta(\log n)$, since
the operator $\text{\texttt{PENC}}_n$ is an extension of
$\text{\texttt{ENC}}^*_n$.

{\bf Summation of bits.} The operator $\text{\texttt{SUM}}_n:
\mathbb B^{n} \to [\mkern-3mu[ n+1 ]\mkern-3mu]$ computes the
arithmetic sum of $n$~boolean variables.

The complexity of bit summation is fairly well studied. Efficient
bit summation circuits are built from compressor subcircuits. A
compressor transforms several bit inputs into a smaller number of
outputs while preserving the sum (taking into account significance
of bits). An example of summation circuit composed of
$(3,\,2)$-compressors $\Sigma_{3,2}$, which compute the sum of
three bits, is shown in Fig.~\ref{pic_cnt}.

\ignore{
\begin{figure}[htb]
\begin{center}
 \includegraphics[scale = 0.36, bb= 0 0 1600 600]{cnt.png}   
 \caption{Стандартная схема суммирования $n$ бит}\label{pic_cnt}
\end{center}
\end{figure}
}


\begin{figure}[htb]
\begin{center}
\begin{picture}(382,144)(-72,0)



\multiput(-45,125)(45,0){5}{\line(0,-1){20}}
\multiput(-45,125)(45,0){5}{\line(1,0){30}}%
\multiput(-15,125)(45,0){5}{\line(0,-1){20}}
\multiput(-45,105)(45,0){5}{\line(1,0){30}}%
\multiput(-39,111)(45,0){5}{$\Sigma_{3,2}$}%

\put(255,125){\line(0,-1){20}}
\put(255,125){\line(1,0){30}}%
\put(285,125){\line(0,-1){20}}
\put(255,105){\line(1,0){30}}%
\put(261,111){$\Sigma_{3,2}$}%

\multiput(-60,115)(45,0){6}{\vector(1,0){15}}
\multiput(240,115)(45,0){2}{\vector(1,0){15}}
\put(203,111){$\cdots$}

\multiput(-40,135)(45,0){5}{\vector(0,-1){10}}%
\multiput(-20,135)(45,0){5}{\vector(0,-1){10}}%
\put(275,135){\vector(0,-1){10}}%


\multiput(22.5,90)(90,0){2}{\line(0,-1){20}}
\multiput(22.5,90)(90,0){2}{\line(1,0){30}}%
\multiput(52.5,90)(90,0){2}{\line(0,-1){20}}
\multiput(22.5,70)(90,0){2}{\line(1,0){30}}%
\multiput(28.5,76)(90,0){2}{$\Sigma_{3,2}$}%

\put(232.5,90){\line(0,-1){20}}
\put(232.5,90){\line(1,0){30}}%
\put(262.5,90){\line(0,-1){20}}
\put(232.5,70){\line(1,0){30}}%
\put(238.5,76){$\Sigma_{3,2}$}%

\multiput(25,105)(90,0){2}{\vector(1,-4){3.75}}
\multiput(50,105)(90,0){2}{\vector(-1,-4){3.75}}
\put(52.5,80){\vector(1,0){60}}
\multiput(142.5,80)(120,0){2}{\vector(1,0){15}}
\put(260,105){\vector(-1,-4){3.75}}

\put(-20,105){\line(1,-2){12.5}} \put(-7.5,80){\vector(1,0){30}}

\put(188,76){$\cdots$}

\multiput(45,70)(90,0){2}{\vector(1,-4){3.75}}
\multiput(50,105)(90,0){2}{\vector(-1,-4){3.75}}
\put(240,70){\vector(-1,-4){3.75}}


\put(121,41){$\cdots$}

\put(112.5,20){\line(0,-1){20}}
\put(112.5,20){\line(1,0){30}}%
\put(142.5,20){\line(0,-1){20}}
\put(112.5,0){\line(1,0){30}}%
\put(118.5,6){$\Sigma_{3,2}$}%

\put(115,35){\vector(1,-4){3.75}}
\put(140,35){\vector(-1,-4){3.75}}
\put(142.5,10){\vector(1,0){20}}


\put(-72,113){$x_1$} \put(-45,137){$x_2$} \put(-25,137){$x_3$}
\put(0,137){$x_4$} \put(45,137){$x_6$} \put(90,137){$x_8$}
\put(133.5,137){$x_{10}$} \put(20,137){$x_5$} \put(65,137){$x_7$}
\put(110,137){$x_9$} \put(153.5,137){$x_{11}$}
\put(270,137){$x_n$}

\put(303,114){$y_0$} \put(280,79){$y_1$}   \put(166,9){$y_{\,\log
n}$}

\end{picture}
\caption{Standard circuit for summing $n$ bits}\label{pic_cnt}
\end{center}
\end{figure}
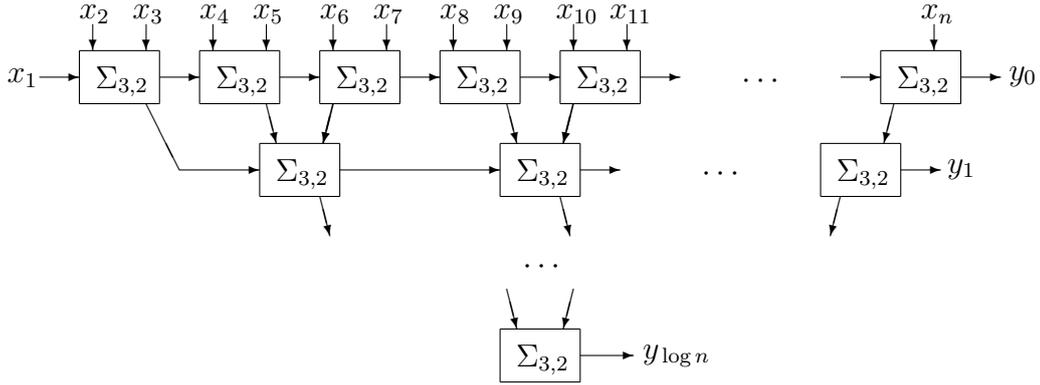


In~\cite{dkky10}, a circuit of complexity ${4.5n-\Theta(\log n)}$
was constructed from special $(5,\,3)$-compressors. At the cost of
increasing the complexity to $4.5n+o(n)$, the circuit can be made
parallel~\cite{serg13}. For this, for example, one can divide the
inputs into groups of $\log n$ pieces, calculate the sums in the
groups by the method~\cite{dkky10}, then sum the group sums via
any parallel compressor circuit. The first stage has complexity
$4.5n-o(n)$, and the second $o(n)$. Thus,
$\mathsf{C}(\text{\texttt{SUM}}_n) \le 4.5n-\Theta(\log n)$ and
$\mathsf{C}_{\log}(\text{\texttt{SUM}}_n) \le 4.5n+o(n)$.
Constructing parallel circuits of compressors is discussed in
detail in~\cite{ppz92}, see also~\cite[\S3.2]{weg87}.

The lower bound can be stated as
$\mathsf{C}(\text{\texttt{SUM}}_n) \ge 2.5n + \Theta(\log n)$. It
is obtained by combining the bound $2.5n-O(1)$ proved by
L.~Stockmeyer~\cite{sto77}, which holds for every component of the
operator $\text{\texttt{SUM}}_n$ except perhaps the lowest and
several highest ones, and the simple relation~\cite{ls73} for the
complexity of a system of functions:
\begin{equation}\label{ls}
  \mathsf{C}(f_1, \ldots, f_k) \ge \min_i \mathsf{C}(f_i) + k-1.
\end{equation}

{\bf Counting the width of a block.} The operator
$\text{\texttt{BW}}_n: \mathbb B^{n} \to [\mkern-3mu[ n+1
]\mkern-3mu]$ for a boolean string of length $n$ that contains a
single block of ones, computes the width of this block.

Upper bounds $\mathsf{C}(\text{\texttt{BW}}_n) \le 4n-\Theta(\log
n)$ and $\mathsf{C}_{\log}(\text{\texttt{BW}}_n) \le 4n+o(n)$ are
proved via the method of compressors. The basic circuit is
constructed as in Fig.~\ref{pic_cnt} from compressor subcircuits
$\Sigma_{3,2}: (a,b,c) \to [u,\,v]$, summing triplets of bits:
$a+b+c = 2u+v$. Provided that the ones at the input are located in
one continuous block, such subcircuits can be implemented with
complexity~4: $u = b(a\vee c),$ $v = a\oplus b \oplus c.$ It is
easy to see that if a string containing a single one-block is fed
to the input of the circuit in Fig.~\ref{pic_cnt}, then the ones
at the input of each internal compressor will be adjacent, in
other words, the input vector $(1,\,0,\,1)$ will never occur. The
total complexity of the described circuit is $4n-\Theta(\log n)$.
A~parallel circuit may be obtained as explained in the previous
paragraph.

{\bf Threshold symmetric functions.} The operator
$\text{\texttt{THR}}^k_n: \mathbb B^{n} \to \mathbb B$ compares
the number of ones in a boolean string $X$ of length $n$ with a
threshold $k$ and thus computes the value of the predicate $\nu(X)
\ge k$.

Upper bounds $\mathsf{C}(\text{\texttt{THR}}_n^k) \le 4.5n+O(\log
n)$ and $\mathsf{C}_{\log}(\text{\texttt{THR}}_n^k) \le 4.5n+o(n)$
follow directly from the corresponding complexity bounds for the
bit summation operator. It suffices to attach a circuit comparing
the sum with the threshold to the circuit implementing
$\text{\texttt{SUM}}_n$. L.~Stockmeyer~\cite{sto77} proved the
lower bound $\mathsf{C}(\text{\texttt{THR}}_n^k) \ge
2n+\min\{k-2,\,n-k-1\}-3$.

Circuits for other symmetric functions, such as periodic ones, can
be constructed similarly.

For a constant threshold $k$, special methods of synthesis allow
to obtain better complexity bounds. In particular,
$\mathsf{C}(\text{\texttt{THR}}_n^2) =
\mathsf{C}_{\log}(\text{\texttt{THR}}_n^2) = 2n+\Theta(\sqrt n)$,
$\mathsf{C}_{\log}(\text{\texttt{THR}}_n^3) \le 3n + O(\log n)$,
and in general for $2^{p-1} < k \le 2^p$ we have
$\mathsf{C}_{\log}(\text{\texttt{THR}}_n^{k}) \le {(4.5 -
2^{2-p})n + \theta_k(n)}$, where $\theta_k(n) = O(\sqrt n)$ or
$\theta_k(n) = O(\log n)$, see~\cite{serg20}.

{\bf Sorting.} The operator $\text{\texttt{SORT}}_n: \mathbb B^{n}
\to \mathbb B^{n}$ sorts a bit string in ascending order (zeros on
the left, ones on the right).

Upper bounds $\mathsf{C}(\text{\texttt{SORT}}_n) \le 6.5n+O(\sqrt
n)$ and $\mathsf{C}_{\log}(\text{\texttt{SORT}}_n) \le 6.5n+o(n)$
are achieved by circuits of the type $\text{\texttt{UN}} \circ
\text{\texttt{SUM}}$. First, the operator $\text{\texttt{SUM}}_n$
calculates the number of ones in the string, then
$\text{\texttt{UN}}_n$ computes the unary representation of this
number.

The lower bound $\mathsf{C}_{\log}(\text{\texttt{SORT}}_n) \ge
3n-6$ is obtained by combining the lower bound $2n-3$ due to
B.~M.~Kloss~\cite{km65}, which holds for any component of the
operator except the lowest and highest (minimum and maximum), and
the bound~(\ref{ls}).

{ \tabcolsep=0.2em \linespread{1.05}
\begin{table}[htb]
\begin{center}
\caption{Complexity bounds for basic operators}\label{t1}
\smallskip
\begin{tabular}{||l|c|c|c||}
\hline
 operator $F$ & \begin{tabular}{c} lower \\ bound $\mathsf{C}(F)$ \end{tabular} &
 \begin{tabular}{c} upper \\ bound $\mathsf{C}(F)$ \end{tabular} & \begin{tabular}{c} upper \\ bound $\mathsf{C}_{\log}(F)$ \end{tabular} \\ \hline
 $\text{\texttt{PREF}}_n$  &  \multicolumn{2}{c|}{ $n-1$} &  $2n-\Theta(\log n)$~\cite{ofm62}  \\ \hline
 $\text{\texttt{PS}}_n$  &  \multicolumn{2}{c|}{ $2n-3$} &  $3n-\Theta(\log n)$  \\ \hline
 $\text{\texttt{INC}}_n$ & \multicolumn{2}{c|}{ $2n-2$ } & $3n-\Theta(\log n)$  \\ \hline
 $\text{\texttt{UDC}}_n$ & --- & $3n-3$ & $4n-\Theta(\log n)$  \\ \hline
 $\text{\texttt{GRC}}_n$ & --- & $4n-7$ & $6n-\Theta(\log n)$  \\ \hline
 $\text{\texttt{CAR}}_n$ & \multicolumn{2}{c|}{ $2n-2$ } & $5n-\Theta(\log n)$ \\ \hline
 $\text{\texttt{ADD}}_n$ & \multicolumn{2}{c|}{ $5n-3$~\cite{red81} } & $8n-\Theta(\log n)$  \\ \hline
 $\text{\texttt{CMP}}_n$ &  --- & $4n-3$ & $5n-\Theta(\log n)$ \\ \hline
 $\text{\texttt{MAX}}_n$ & --- & $6n-3$ & $7n-\Theta(\log n)$   \\ \hline
 $\text{\texttt{DEC}}_n$  &  \multicolumn{3}{c||}{ $n + \Theta(\sqrt n)$ } \\ \hline
 $\text{\texttt{MUX}}_n$ & $2n-2$~\cite{paul77} &   \multicolumn{2}{c||}{ $2n+O(\sqrt n)$~\cite{kp80} } \\ \hline
 $\text{\texttt{MUX}}^k_n$ & --- &   \multicolumn{2}{c||}{ $2kn+O(\sqrt {kn})$ } \\ \hline
 $\text{\texttt{CYC}}_{k,n}$ & --- & \multicolumn{2}{c||}{ $3\lceil \log k\rceil n$ }  \\ \hline
 $\text{\texttt{SFT}}_{k,n}$ & --- & \multicolumn{2}{c||}{ $3\lceil \log k\rceil n - \Theta(k)$ }  \\ \hline
 $\text{\texttt{ENC}}_n$  &  \multicolumn{3}{c||}{ $2(n-\lceil \log n\rceil -1)$~\cite{cha94} } \\ \hline
 $\text{\texttt{UN}}_n$  & --- & \multicolumn{2}{c||}{ $2n+O(\sqrt n)$ } \\ \hline
 $\text{\texttt{UN}}^{-1}_n$  & \multicolumn{3}{c||}{ $n-1$ } \\ \hline
 $\text{\texttt{TRN}}_n$  & --- & \multicolumn{2}{c||}{ $3n+O(\sqrt n)$ } \\ \hline
 $\text{\texttt{FOI}}_n$  & \multicolumn{2}{c|}{ $2n-2$ } & $3n-\Theta(\log n)$ \\ \hline
 $\text{\texttt{PENC}}_n$  &  $2n-\Theta(\log n)$ & $2n-3$ & $3n-\Theta(\log n)$  \\ \hline
 $\text{\texttt{SUM}}_n$ & $2.5n+\Theta(\log n)$~\cite{sto77,ls73} & $4.5n-\Theta(\log n)$~\cite{dkky10}  &  $4.5n+o(n)$~\cite{serg13}  \\ \hline
 $\text{\texttt{THR}}_n^k$ & $2n+\min\{k,\,n-k\}-5$~\cite{sto77} & $4.5n+O(\log n)$ & $4.5n+o(n)$ \\ \hline
 $\text{\texttt{BW}}_n$ & --- & $4n-\Theta(\log n)$ &  $4n+o(n)$  \\ \hline
 $\text{\texttt{SORT}}_n$ & $3n-6$~\cite{km65,ls73} & $6.5n+O(\sqrt n)$ & $6.5n+o(n)$  \\ \hline
\end{tabular}
\end{center}
\end{table}
}

\medskip


{\bf \qquad Some applications}

\medskip

In this section, we discuss several applications of the
constructions mentioned above to building some specific parallel
circuits. The first example illustrates the use of parallel
prefix-suffix circuits. The second example relies on an efficient
implementation of unary-binary transforms. The third example
illustrates the principle of mass production and leads to the
fourth example.

{\bf Two-selector.} The operator $\text{\texttt{TOI}}_n: \mathbb
B^{n} \to \mathbb B^{n}\times \mathbb B$ preserves two ones in a
boolean string of length $n$, replacing the rest with zeros, and
additionally computes the indicator of the presence of ones in the
string. This operator is an extension of the operator
$\text{\texttt{FOI}}_n$ and is used to select two active channels
marked with ones in a set of $n$ data channels, see,
e.g.,~\cite{ajy04}.

The bound $\mathsf{C}_{\log}(\text{\texttt{TOI}}_n) \le 5n -
\Theta(\log n)$ is achieved by a circuit that selects two extreme
ones on different sides. Given the original string $[x_1, \ldots,
x_n]$, compute the string of prefix sums $[p_1, \ldots, p_n] =
[0]^k[1]^{n-k}$ and the string of suffix sums $[s_1, \ldots, s_n]
= [1]^l[0]^{n-l}$ applying the operator
$\text{\texttt{PS}}_n^{\vee}$, where $k+1$ and $l$ are the
positions of the first and last ones (if there are ones). In this
case, $p_n=s_1$ serves as an indicator of the presence of ones.
The resulting string $[z_1, \ldots, z_n]$ may be computed bitwise
as $z_i = x_i \cdot ( \overline{p_{i-1}} \vee
\overline{s_{i+1}})$.

{\bf Weight-preserving counter}. The operator
$\text{\texttt{NCK}}_n: [\mkern-3mu[ 2^n ]\mkern-3mu] \to
[\mkern-3mu[ 2^n ]\mkern-3mu]$ computes the next number in
ascending order after a given one with the same binary weight. If
there is no greater number, the output is the same as the input.
This operator was discussed, in particular, in~\cite{ny05} in
connection with applications in image processing. There it is
called $\binom{n}k$-counter. The complexity of a parallel
implementation of the operator $\text{\texttt{NCK}}_n$
in~\cite{ny05} is stated as $O(n)$, but a bound of about~$16n$ may
be extracted from the circuit description.

The upper complexity bound can be refined to
$\mathsf{C}_{\log}(\text{\texttt{NCK}}_n) \le 13n+o(n)$. By
appending an extra zero to the left of a non-zero $n$-bit number,
this number can be uniquely represented as $S \,\|\,
0\,[1]^j[0]^i$, where $i \ge 0$, $j \ge 1$, and $S$ is a non-empty
substring in the case when the number is not maximal among numbers
of the same weight. Then the operator $\text{\texttt{NCK}}_n$
transforms this non-maximal number as $S \,\|\, 0\,[1]^j[0]^i \to
S \,\|\, 1\,[0]^{i+1}[1]^{j-1}$. The sequence of computations
leading to the required result is shown in Fig.~\ref{pic_nck},
where $*$ denotes irrelevant parts of strings, \texttt{bit[]}
denotes the corresponding bitwise operation. The arguments and
results of some operations are shifted by 1. Extra bits are shown
for generality of notation; they are not used in the calculations.
The circuit consists of subcircuits implementing the operators
$\text{\texttt{PREF}}_n^{\vee}$, $\text{\texttt{UN}}_n$,
$\text{\texttt{UN}}_n^{-1}$ (twice), seven bitwise operations with
$n$-bit vectors, and subtraction of $\log n$-bit numbers. The
indicator $c$ of the maximality of an input number can be written
as $(i+j = n)$. It is computed in the second step as the second
from the left (i.e., the most significant) bit of the number
$[0]^{n+1-i-j}[1]^{i+j}$. In the final step of the algorithm, when
$c=1$, the input number is selected; when $c=0$, the result is
composed of the fragment~$S$ of the input number and the computed
fragment $1\,[0]^{i+1}[1]^{j-1}$. The circuit also works correctly
with zero input.

\ignore{
\begin{figure}[t]
\begin{center}
 \includegraphics[scale = 0.4, bb= 0 0 1735 465]{nck.png}   
 \caption{Диаграмма алгоритма вычислений в схеме счетчика с
сохранением веса}\label{pic_nck}
\end{center}
\end{figure}
}


\begin{figure}[htb]
\begin{center}
\begin{picture}(417,110)(-49,-14.5)

\thicklines
\put(28,53.5){\oval(64,16)} 
\put(-6,-1){\oval(86,27)} 
\thinlines
\put(-2,50){$S \,\|\, 0\,[1]^j[0]^i$} 
\put(134.5,53.5){\oval(56,16)} 
\put(108.5,50){$* \,\|\, 1\,[0]^{i+j}$} 
\put(243,50.5){\oval(85,24)} 
\put(205.5,50){$[0]^{n+1-i-j}[1]^{i+j},$} 
\put(240,41){$c$} 
\put(291,83.5){\oval(63,16)} 
\put(264,80){$[0]^{n+1-i}[1]^{i}$} 
\put(323,50){\oval(29,13)} 
\put(311,47){$i+j$} 
\put(362,51){\oval(12,12)} 
\put(360,47.5){$i$} 
\put(292.5,2.5){\oval(12,14)} 
\put(290,0){$j$} 
\put(227,3.5){\oval(74,16)} 
\put(195,0){$[0]^{n+2-j}[1]^{j-1}$} 
\put(103,3.5){\oval(82,16)} 
\put(63.5,0){$* \,\|\, 1\,[0]^{i+1}[1]^{j-1}$} 
\put(-47,0){$S \,\|\, 1\,[0]^{i+1}[1]^{j-1},$} 
\put(-28,-13){if $c = 0$} 

\put(61,55){\vector(1,0){44}} 
\put(61,51){\vector(1,0){44}} 
\put(24,43){\vector(1,-3){11}} 
\put(164,53){\vector(1,0){35}} 
\put(143,43){\vector(0,-1){32}} 
\put(287,50){\vector(1,0){20}} 
\put(199,47){\vector(-4,-1){161}} 
\put(320,42){\vector(-2,-3){22}} 
\put(355,48){\vector(-4,-3){55}} 
\put(285,3){\vector(-1,0){20}} 
\put(61,3){\vector(-1,0){22}} 
\put(189,3){\vector(-1,0){43}} 
\put(30,63){\line(2,1){40}} 
\put(70,83){\vector(1,0){188}} 
\put(242,64){\vector(1,1){16}} 
\put(322,78){\vector(3,-2){34}} 

\put(63,59){\small \texttt{bit[$\scriptstyle{\overline{x}\cdot y}$]}}  
\put(170,57){\small $\overline{\text{\texttt{PREF}}^{\vee}}$}  
\put(217,88){\small \texttt{bit[$\scriptstyle{\overline{x} \cdot y}$]}}  
\put(287,54){\small $\text{\texttt{UN}}^{-1}$}  
\put(338,70){\small $\text{\texttt{UN}}^{-1}$}  
\put(271,8){\small $\text{\texttt{UN}}$}  
\put(147,8){\small \texttt{bit[$\scriptstyle{x\vee y}$]}}  
\put(30.5,26.5){\small \texttt{bit[$\scriptstyle{(y\vee c)(x \oplus z) \oplus z}$]}}  

\end{picture}
\caption{Computational scheme of the weight-preserving counting
circuit}\label{pic_nck}
\end{center}
\end{figure}
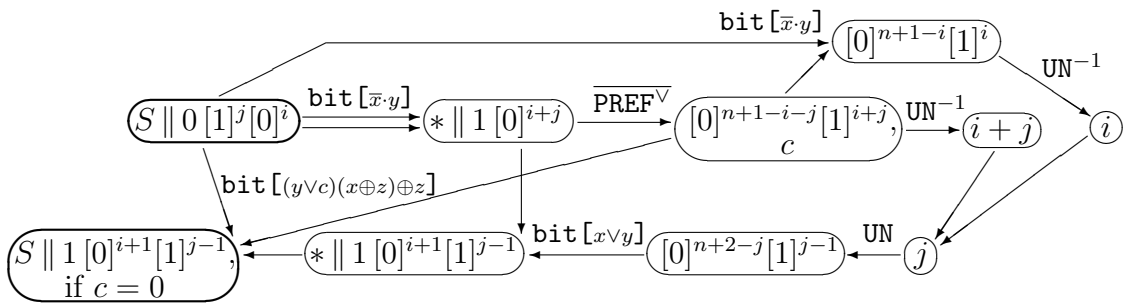


In practical circuit design, the condition $\overline c$ can be
used separately as an indicator of the result's validity (for
storing it in a register), then the final operation can be
simplified to \texttt{bit[$\scriptstyle{xy\vee z\overline y}$]}.
In this case, the circuit complexity decreases to $12n+o(n)$.

{\bf Multiple selection.} Let us consider another generalization
of the multiplexor: the operator $\text{\texttt{SEL}}^k_n:
[\mkern-3mu[ n ]\mkern-3mu]^k \times \mathbb B^{n} \to \mathbb
B^k$ selects $k$ of $n$ boolean information variables by their
addresses. D.~Uhlig's method (see~\cite{uhl92,weg87}) allows to
construct smaller circuits for multiselection compared to a
straightforward collection of $k$ independent multiplexor
circuits. To demonstrate the idea, we restrict ourselves to the
case $k=2$.

Let us show that $\mathsf{C}_{\log}(\text{\texttt{SEL}}^2_n) \le
3n + O(n^{2/3})$. Denote $m=\lceil \log n \rceil$. Split a string
$Y \in \mathbb B^n$ into $2^r$ substrings $Y_i$, $0 \le i < 2^r$,
where the string $Y_i = [y_i,\,y_{i+2^r},\,y_{i+2\cdot2^r},
\ldots]$ includes variables with indices equal to $i$ modulo
$2^r$. By construction, $|Y_i| = q := \lceil n/2^{r} \rceil$ (we
pad shorter strings to length~$q$ arbitrarily). Compute the
strings
\begin{equation}\label{wtY}
\widetilde Y_0 = Y_0, \;\, \widetilde Y_1 = Y_0 \oplus Y_1, \;\, \ldots, \;\, \widetilde Y_{i} = Y_{i-1} \oplus Y_{i}, \;\, \ldots, \;\, 
\widetilde Y_{2^r-1} = Y_{2^r-2}\oplus Y_{2^r-1}, \;\, \widetilde
Y_{2^r} = Y_{2^r-1},
\end{equation}
where addition operations are bitwise. Note that for any $i$,
\[ \widetilde Y_0 \oplus \ldots \oplus \widetilde Y_{i} = Y_i = \widetilde Y_{i+1} \oplus \ldots \oplus \widetilde Y_{2^r}. \]

Let $a = [a_1,\, a_0]$ and $b = [b_1,\, b_0]$ denote the address
inputs of the circuit -- there we distinguish groups of the least
significant $r$ bits: $|a_0|=|b_0|=r$.

Thus, if $a_0 \le b_0$, then we can determine
\[ \text{\texttt{SEL}}^2_n(a, b; Y) = [\text{\texttt{MUX}}_n(a; Y),\, \text{\texttt{MUX}}_n(b; Y)] = [\text{\texttt{MUX}}_q(a_1; Y_{a_0}),\,\text{\texttt{MUX}}_q(b_1; Y_{b_0})] \]
implementing operators $\text{\texttt{MUX}}_q(z_i; \widetilde
Y_i)$, while setting $z_i=a_1$ for $0 \le i \le a_0$ and $z_i=b_1$
for ${b_0 < i \le 2^r}$. In the case $a_0 > b_0$, the roles of $a$
and $b$ are swapped. In accordance with this rule, we introduce
indicator functions
\begin{equation}\label{eta}
\eta^a_i = (i \le a_0 \le b_0) \vee (i > a_0 > b_0), \quad
\eta^b_i = (i > b_0 \ge a_0) \vee (i \le b_0 < a_0),
\end{equation}
which choose whether to use the subcircuit
$\text{\texttt{MUX}}_q(z_i; \widetilde Y_i)$ to compute
$\text{\texttt{MUX}}_n(a; Y)$ or to compute
$\text{\texttt{MUX}}_n(b; Y)$. The final result is determined by
formulas
\begin{multline}\label{sel}
\text{\texttt{MUX}}_n(a; Y) = \bigoplus_{i=0}^{2^r} \eta^a_i\, \text{\texttt{MUX}}_q(z_i; \widetilde Y_i), \quad \text{\texttt{MUX}}_n(b; Y) = \bigoplus_{i=0}^{2^r} \eta^b_i\, \text{\texttt{MUX}}_q(z_i; \widetilde Y_i), \\
z_i = [\eta^a_i]^{m-r} \cdot a_1 \vee [\eta^b_i]^{m-r} \cdot b_1,
\end{multline}
where the operations in the last formula are bitwise. The circuit
complexity results from the computations implied
by~(\ref{wtY}),~(\ref{eta}),~(\ref{sel}), and is estimated as
\begin{equation*}
 \mathsf{C}_{\log}(\text{\texttt{SEL}}_n^2) \le (2^r-1)q + (2^r+1)\mathsf{C}_{\log}(\text{\texttt{MUX}}_q) +  O(2^rm) \le 3\cdot2^rq + O(2^r(m+\sqrt q) + q).
\end{equation*}
The required bound is obtained by choosing $r \approx \log n / 3$.
The method is already practical for $r=1$.

In fact, a linear complexity upper bound
$\mathsf{C}_{\log}(\text{\texttt{SEL}}^k_n) = O(n)$ holds for all
$k \le n/\log^3 n$, as shown in~\cite{hr24} by a rather nontrivial
method.

{\bf Permutation of a pair of bits.} The operator
$\text{\texttt{EXC}}_n: [\mkern-3mu[ n ]\mkern-3mu]^2 \times
\mathbb B^{n} \to \mathbb B^n$ permutes two bits with given
addresses in a boolean string of length $n$. This fairly popular
operation (especially in programming) is discussed, for example,
in~\cite[\S7.1.3]{knu4}.

Based on the result of the previous paragraph, it is easy to
derive the bound $\mathsf{C}_{\log}(\text{\texttt{EXC}}_n) \le 7n
+ O(n^{2/3})$. The solution also exploits the well-known trick of
exchanging the contents of two registers,
see~\cite[Chapter~2]{war03}. First, compute the required pair of
bits $[u,\, v] = \text{\texttt{SEL}}^2_n(a,b;Y)$. Then, via two
demultiplexors, produce the strings ${\text{\texttt{DEC}}^*_n(a; u
\oplus v)}$ and ${\text{\texttt{DEC}}^*_n(b; u \oplus v)}$
containing bits ${u \oplus v}$ in each of the two given positions
$a, b$. At last, add these strings bitwise to each other and to
the input string $Y$ (modulo $2$).

\medskip

The author is grateful to a referee for useful comments, in
particular for observations that allowed to refine the complexity
bounds of the conversion from unary to binary encoding, the
priority encoder, and the two-selector.

\end{document}